\documentclass[12pt]{article}
\usepackage[style=numeric-comp,giveninits,maxnames=6,doi=false]{biblatex}
\bibliography{paper19-arxiv-2023-v1.bib}
\usepackage[american]{babel}
\usepackage{csquotes,mathtools,setspace,authblk,ebgaramond-maths}
\usepackage[colorlinks,allcolors=blue]{hyperref}

\title{\bf Nomic realism, simplicity, and the simplicity bubble effect}

\author[1,2,3,4]{Felipe S. Abrahão\thanks{E-mail: \href{mailto:fabrahao@unicamp.br}{fabrahao@unicamp.br}.}}
\author[2,5]{Raoni Arroyo\thanks{E-mail: \href{mailto:rwarroyo@unicamp.br}{rwarroyo@unicamp.br}, corresponding author.}}

\affil[1]{\small Oxford Immune Algorithmics, Oxford University Innovation.\newline University of Oxford. UK.}
\affil[2]{\small Centre for Logic, Epistemology and the History of Science.\newline University of Campinas. Brazil.}
\affil[3]{\small Data Extreme Lab, National Laboratory for Scientific Computing. Brazil.}
\affil[4]{\small Algorithmic Nature Group, Laboratoire de Recherche Scientifique (LABORES)\newline for the Natural and Digital Sciences. France.}
\affil[5]{\small Department of Philosophy, Communication, and Performing Arts.\newline Roma Tre University. Italy.}

\date{Contributed talk for the Third Graduate Conference of the Italian Network for the Philosophy of Mathematics --- FilMat. Submitted: September 15, 2023. Approved: October 25, 2023.}

\begin{document}
\sloppy\raggedbottom
\maketitle\onehalfspacing

\begin{abstract}
    We offer an argument against simplicity as a sole intrinsic criterion for nomic realism. The argument is based on the simplicity bubble effect. Underdetermination in quantum foundations illustrates the case.
\end{abstract}

\section{Nomic underdetermination, the quantum case}
Nomic realists say the world has nomic structure, \textit{i.e.}, there are laws of Nature. Here's a puzzle for them: according to quantum mechanics, what are such laws? Due to underdetermination in responses to the measurement problem, the answer cannot be straightforward: we have as many candidates for laws of quantum mechanics as there are interpretations of quantum mechanics.

The well-known measurement problem consists of leveling what the formalism of quantum mechanics is telling us with what laboratory outcomes are telling us. Given a specific basis, quantum mechanics mathematically describes physical objects by means of wave functions which might be, in turn, described in terms of a superposition of multiple quantum states in the form of, \textit{e.g.}, $\sum c_i | \alpha_i \rangle$. When in contact with a calibrated measuring apparatus, $|M_0 \rangle$, quantum mechanics predicts that system and apparatus will become correlated so that the composed system $\langle\text{quantum system}+\text{measuring apparatus}\rangle$ will be described by $\sum c_i | \alpha_i \rangle |M_i \rangle$. As a consequence of such dynamics, the wave function of the measuring apparatus (\textit{e.g.}, its pointers) has also become a superposition $\sum c_i |M_i \rangle$. The odd thing it is hard to appreciate the physical meaning of superposed macroscopic states, such as pointers of measuring apparatuses. This is the measurement problem in a nutshell, and to solve it is to offer an interpretation of quantum mechanics.

Numerous solutions have been proposed; as we need only a pair of them in order to build an underdetermination argument, we'll reduce the sampling to two. One of them is to postulate that due to its increasing complexity, wave functions spontaneously collapse into one of its components in the form of $\sum c_i | \alpha_i \rangle |M_i \rangle\implies c_k | \alpha_k \rangle |M_k \rangle_{k\in{i}}$. This requires one to modify the standard quantum formalism, supplementing it with two constants of nature, \textit{viz.} the collapse frequency and amplitude \cite{grw1986}. The second approach consists of taking superpositions literally, in which every term of a superposition actually occurs in a different ``world'', so every time quantum systems interact with measuring apparatuses reality branches into different worlds (as many as the terms of the superposition, \textit{e.g.}, the range of the index $i$) \cite{everett1957}.

On the nomic realist side, it might be argued that wave function collapse is a law of nature; alternatively, one might argue that it isn't but branching is instead. Both options are empirically adequate with regard to current empirical evidence---they're currently underdetermined by data---, so the debate of which one is the \textit{true} one cannot happen in the empirical arena. 

\section{Simplicity as a tie-breaker}
This is why extra-empirical criteria have been put forth to decide between them. Among such criteria is the criterion of simplicity: we should adopt the \textit{simpler} alternative, as simplicity is an explanatory virtue (epistemic, pragmatic, or both). Although simplicity has been under attack for many years for not being truth-conducive---after all, Nature is under no obligation of being simple---recently the debate has changed: instead of being \textit{truth}-conducive, it has been argued that simplicity is \textit{law}-conducive, \textit{i.e.} we are more likely to grasp simple laws of nature \cite{chen-goldstein2022,chen2023,chen2023laws}. So even though this is not an argument for the scientific realist stance, it is an argument in favor of the nomic realist stance. This would arguably favor versions of Everettian quantum mechanics, as it requires a much simpler structure than its alternative rivals (\textit{e.g.} Bohmian mechanics and GRW): it is just linear quantum mechanics by itself, without the troublesome wave function collapse \cite{wallace2012}.

Nomic simplicity has presented a way out of the puzzle. In this talk, we'll resist that by presenting an objection to simplicity as a sole ally (or dominant conductive factor) for the nomic realist.

\section{Against `simplicity'$\text{?}$}
Well-known both in practice and in the theoretical foundations of artificial intelligence systems, the problem of overfitting---\textit{viz.}, the error on the test set is considered to be higher than the error on the training set---in machine learning (ML) is usually tackled by employing some heuristic that enforces a bias towards simplicity when looking for the optimal model candidates \cite{Goodfellow2016}.
Such a bias effects the principles of Parsimony, or Occam's razor, in the form of mathematical constraints that induce the computational analysis to favor models that can be generalized beyond the data upon which the learning algorithm was trained, \textit{e.g.}, as done in regularization \cite{Goodfellow2016}.
This generalization compromise from available data assumes a more fundamental instantiation in the problem of local \textit{versus} global optima.
That is, the problem of guaranteeing that the optimal solution found is not a mere local one (or at least, that the learner can avoid being trapped into local optima) but also a global optimum that models all possible data, and not only the data previously made available.

Overcoming this kind of problem faces theoretical limitations and evinces the role of such a bias toward simpler solutions or theories both in artificial intelligence and data science in general.
First, when averaging over all possible ways to generate data, we know from the no free lunch theorem \cite{Wolpert1997} that no single learning algorithm is universally any better than any other learning algorithm.
Applying a bias toward simplicity by constraining the possible ways that the datasets of interest can be generated (for example with additional assumptions regarding the behavior of data in real-world scenarios) is an approach to circumvent this problem in ML. 
Secondly, although computational analysis benefits from larger datasets that are accurate and fine grained, the larger the dataset the more likely it becomes for the data mining algorithms to discover (or uncover) spurious correlations \cite{Calude2017,Smith2020}.
In practice, the data scientist often circumvents this issue by previously selecting a feature or pattern of interest that should be mined from the datasets, which in turn constrains the scope of the observer (i.e., the data analyst) by enforcing the above bias toward simplicity in an a priori manner.

ML techniques have shown a considerable amount of success in dealing with these problems for large amounts of data in real-world scenarios.
However, it can be demonstrated that for a particular type of learner and loss function metric, there is an infinite class of problems for which the optimal solution exists but no learning algorithm can compute it \cite{Colbrook2022}.
In this way, the immediate question that arises is whether or not such an impossibility result can be shown to be fundamental to ML in general.

Indeed, recent studies has shown that in general, even if one assumes the universe of possible data can be ultimately explained by formal mathematical theories, no learning algorithm (with or without the help of an arbitrarily chosen formal theory) can guarantee with arbitrarily low probability that it is calculating the global optimum, and not the local one \cite{Abrahao2021dSBpaperarxivBEPE}.
Ironically, the bias toward simplicity is the main ``driving force'' underlying this mathematical phenomenon so that if the data made available amounts to a sufficiently large quantity, then there is a non-negligible probability that the arbitrarily chosen learning algorithm will be deceived into thinking it found the globally optimal solution while it did in fact find a local one.
This is called the ``simplicity bubble effect'' \cite{Abrahao2021dSBpaperarxivBEPE,Abrahao2023SBZemblanityarxivBEPE}.

\section{A working hypothesis}
Due to this effect, simplicity alone cannot give epistemic warrants to formal theories.
A formal theory that explains (or is always corroborated by) a sufficiently large set of relevant empirical phenomena can never be assured to keep doing so with the same degree of success in the long run.
And this holds even in the case every possible empirical phenomenon can in principle be explained or generated by a hypothetical formal theory.
In order to avoid this deceiving fundamental limitation, additional conditions or assumptions on the relationship between our formal knowledge and the universe's data-generating processes shall take place beyond (or in addition to) the bias toward simplicity.
If this deceiving effect can be extended to the nomic level, then simplicity alone (as an intrinsic law-conductive property of our theories or our empirically verified and justified formal knowledge) \textit{also} cannot give the epistemic infrastructure the nomic realist wants for the exact same reason. 
This is our working hypothesis.

\section*{Acknowledgments}

The authors acknowledge the support from São Paulo Research Foundation (FAPESP): grant \# $2023$/$05593$-$1$ (Felipe S. Abrahão); and grant \# $2022$/$15992$-$8$ (Raoni Arroyo).

\printbibliography
\end{document}